\documentclass[doublecol]{epl2}

\usepackage{amsmath}
\usepackage{amssymb}
\usepackage{graphicx}
\usepackage{color}

\title{Degree of randomness: numerical experiments for astrophysical signals}
\shorttitle{Degree of randomness: numerical experiments for astrophysical signals}

\author{V.G.Gurzadyan \and T.Ghahramanyan \and S.Sargsyan 
   }

\shortauthor{V.G.Gurzadyan \etal}

\institute{
Alikhanian National Laboratory and Yerevan State University, Yerevan, Armenia
}

\pacs{98.80.-k}{Cosmology}
\pacs{98.70.Vc}{Background radiations}

\abstract{Astrophysical and cosmological signals such as the cosmic microwave background radiation, as observed, typically contain contributions
of different components, and their statistical properties can be used to distinguish one from
the other. A method developed originally by Kolmogorov is involved for the study of astrophysical signals of randomness of various degrees. Numerical 
performed experiments based on the universality of Kolmogorov distribution and using a single scaling of the ratio of stochastic to regular 
components, reveal basic features in the behavior of generated signals also in terms of a critical value for that ratio, thus enable the 
application of this technique for various observational datasets.    
}

\begin{document}

\maketitle



\section{Introduction}

Kolmogorov stochasticity parameter approach has been used to quantify the degree of randomness of sequences of number theory or dynamical 
systems \cite{Kolm,K1940,Arnold}. The stochasticity degree of the fractional parts of the arithmetical progressions has been analysed, 
also in comparison with geometrical progressions\cite{Arnold,Arnold_ICTP,Arnold_UMN,Arnold_MMS,Arnold_FA}. 

The crucial aspect of the application of the method is the behavior of the stochasticity parameter. Arnold studied the randomness for 
arithmetical progressions of the residues for the division by a real number and other sequences, using the  uniform or the Legendre-Chebyshev 
distributions \cite{Arnold_UMN,Arnold_MMS}, mentioning also the problems for which no solutions are known yet.  

Along with this, physical applications of this descriptor of the degree of randomness have been also undertaken. Namely, astrophysical signals 
which contain several subsignals, i.e. contributions of different physical mechanisms, are of particular interest. Cosmic microwave 
background (CMB) radiation, as observed, does contain besides the cosmological signal, also contributions of Galactic and other origins 
\cite{CMB1,CMB2}. It appears that the different degree of randomness quantified by Kolmogorov distribution enables to distinguish the cosmological 
and non-cosmological signals \cite{GK_KSP,G2009}. The fractions of the random and correlated Gaussian components in the CMB overall signal have 
been obtained using that method. It was also efficient for the identification in the CMB maps obtained by the Wilkinson Microwave Anisotropy 
Probe \cite{J} of point sources \cite{G2010}, i.e. of radio, as well as gamma-ray sources observed by the Fermi satellite.

The properties of the CMB signal follow from primordial fluctuations which appear to be close to Gaussian ones. This remarkable empirical fact 
is fortunate for our approach, since it unambiguously 
defines the theoretical cumulative distribution function required to compute the stochasticity parameter. 
Although certain non-Gaussian features have been observed in the CMB datasets (see \cite{spot,Copi,G2,G_plane}) and they are studied by various 
descriptors, the Gaussianity remains as one of the robust characteristics of CMB.  However in other astrophysical signals, for example, for 
Lyman-alpha forest, the mentioned distribution function cannot be so well defined and therefore the problem of statistics becomes more complex. 
In such conditions, the numerical experiments are the means to explore the behavior of the stochasticity parameter vs the generated random and 
regular components of the signal. Thus the universality of Kolmogorov's theorem is elaborated in the numerical experiments.                   

\section{Kolmogorov statistic}

First, let us briefly review the method.  
Consider $\{X_1,X_2,\dots,X_n\}$ independent values of the same real-valued random variable $X$ in growing order  $X_1\le X_2\le\dots\le X_n$ 
and let \cite{Kolm,Arnold}
\begin{equation}
F(x) = P\{X\le x\}\ 
\end{equation}
be a cumulative distribution function of $X$. Their empirical distribution function $F_n(x)$ is defined as

\begin{equation}
F_n(X)= \left\{
\begin{array}{rl}
	0, & X < x_1 \\
	k / n, & x_k \leq X < x_{k+1} \\
	1, & x_n \leq X\\
\end{array}
\right.
\label{eq:empiricdistribution}
\end{equation}Kolmogorov's stochasticity parameter is 
\begin{equation}\label{KSP}
\lambda_n=\sqrt{n}\ \sup_x|F_n(x)-F(x)|\ .
\end{equation}

Kolmogorov theorem \cite{Kolm} states that for any continuous $F$
\begin{equation}
\lim_{n\to\infty}P\{\lambda_n\le\lambda\}=\Phi(\lambda)\ ,
\end{equation}
where $\Phi(0)=0$,
\begin{equation}
\Phi(\lambda)=\sum_{k=-\infty}^{+\infty}\ (-1)^k\ e^{-2k^2\lambda^2}\ ,\ \  \lambda>0\ ,\label{Phi}
\end{equation}
the convergence is uniform and Kolmogorov's distribution $\Phi$ is independent on $F$.
 
Since the stochasticity parameter itself is a random quantity its probable values are defined by the distribution function $\Phi$, namely, 
the interval $0.3\le\lambda_n\le 2.4$, where $\lambda$ provides information on the degree of randomness. 
The sky map of the degree of randomness for CMB is shown in Fig. 1: the map contains both the cosmological, as well as non-cosmological 
signals, e.g. the clearly visible Galactic disk (for details see \cite{G2009}). Below we will study more tiny appearances of the Kolmogorov 
function vs the properties of the signals.

\begin{figure}%
\includegraphics[width=3.0 in]{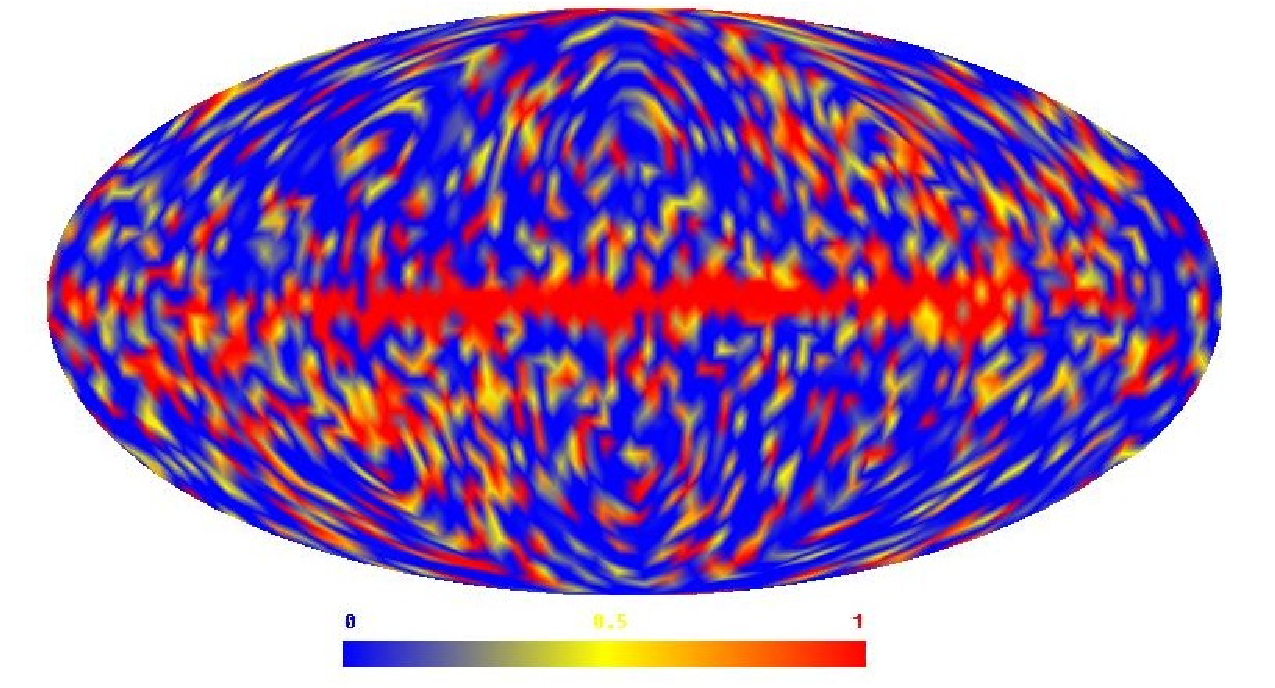}
\caption{The distribution of Kolmogorov's function $\Phi$ for the full sky cosmic microwave background temperature data obtained by 
Wilkinson Microwave Anisotropy Probe.}%
\end{figure}

\section{Stochastic vs regular components}

Using the Kolmogorov method we will study the properties of sequences of type 
\begin{equation}
z_n = \alpha x_n + (1-\alpha) y_n,
\end{equation}
where $x_n$ are random sequences and 
\begin{equation}
y_n = \frac{an\pmod b}{b},
\end{equation}
are regular sequences, $a$ and $b$ are mutually fixed prime numbers, both sequences within the interval $(0,1)$ and have uniform distribution, $\alpha$ indicating the fraction of random and regular sequences. This representation is a more general case of random and regular sequences considered in \cite{mod,G2009}. Other techniques for dealing with astrophysical systems with random (chaotic) and regular behavior are discussed in \cite{GP}.    

When doing statistic with large number of sequences, each new sequence $y_n$ is taken as the continuation of the previous one from the same arithmetical progression.

Thus we have $z_n$ with a distribution function 

\begin{equation}
F(X)= \left\{
\begin{array}{rl}
	0, & X \leq 0 \\
	\frac{X^2}{2 \alpha (1-\alpha)}, & 0 < X \leq \alpha\\
	\frac{2 \alpha X - \alpha^2}{2 \alpha (1-\alpha)}, & \alpha < X \leq 1-\alpha\\
	1-\frac{(1-X)^2}{2 \alpha (1-\alpha)}, & 1-\alpha < X \leq 1\\
	1, & X > 1.\\
\end{array}
\right.
\label{eq:d}
\end{equation}

We will analyze the stochastic properties of $z_n$ vs the parameter $\alpha$ varying between $0$ and $1$ for different values of $a$ and $b$, i.e. 
corresponding to from purely stochastic to purely regular sequences. 

\begin{figure}%
\includegraphics[width=2.37 in]{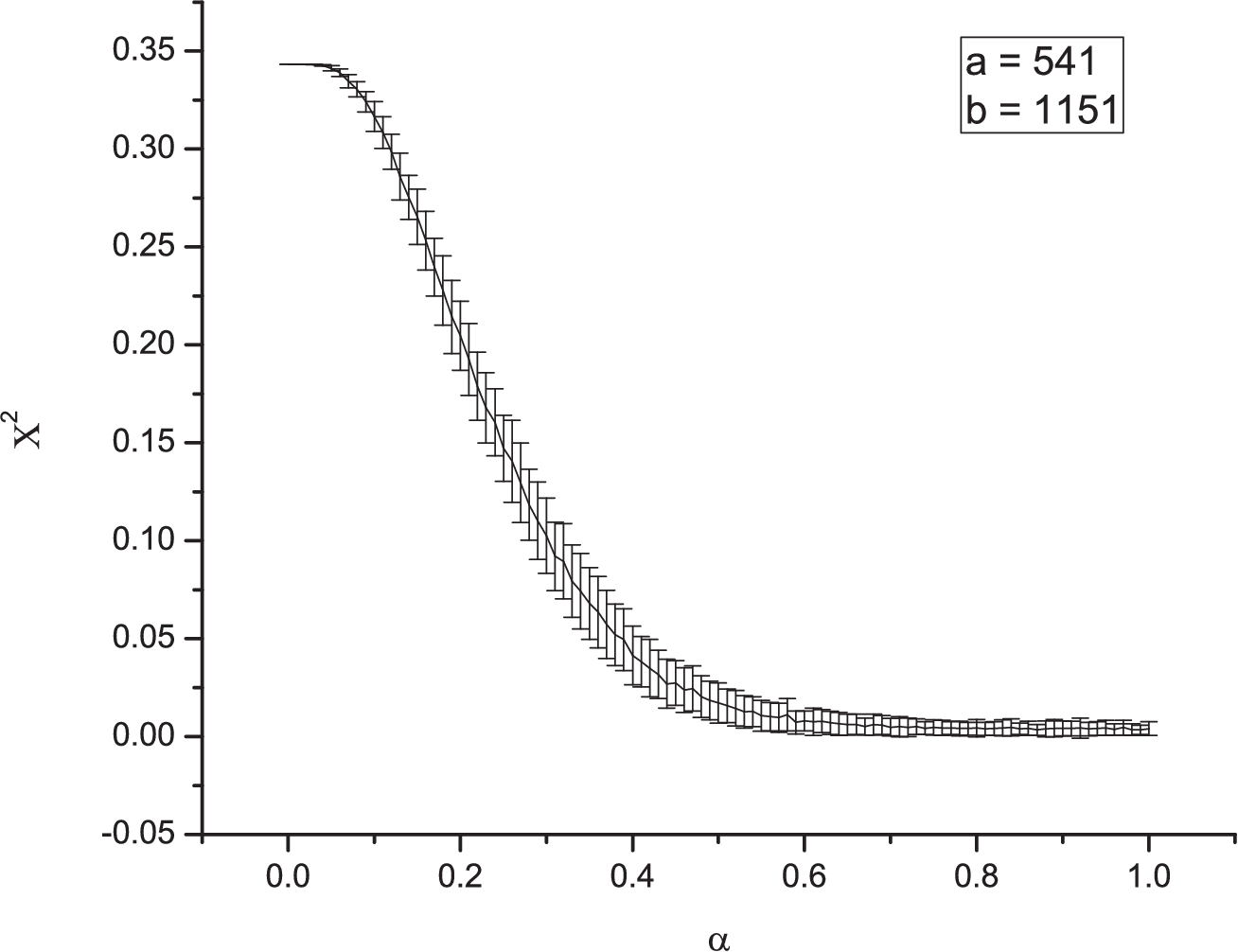}
\includegraphics[width=2.37in]{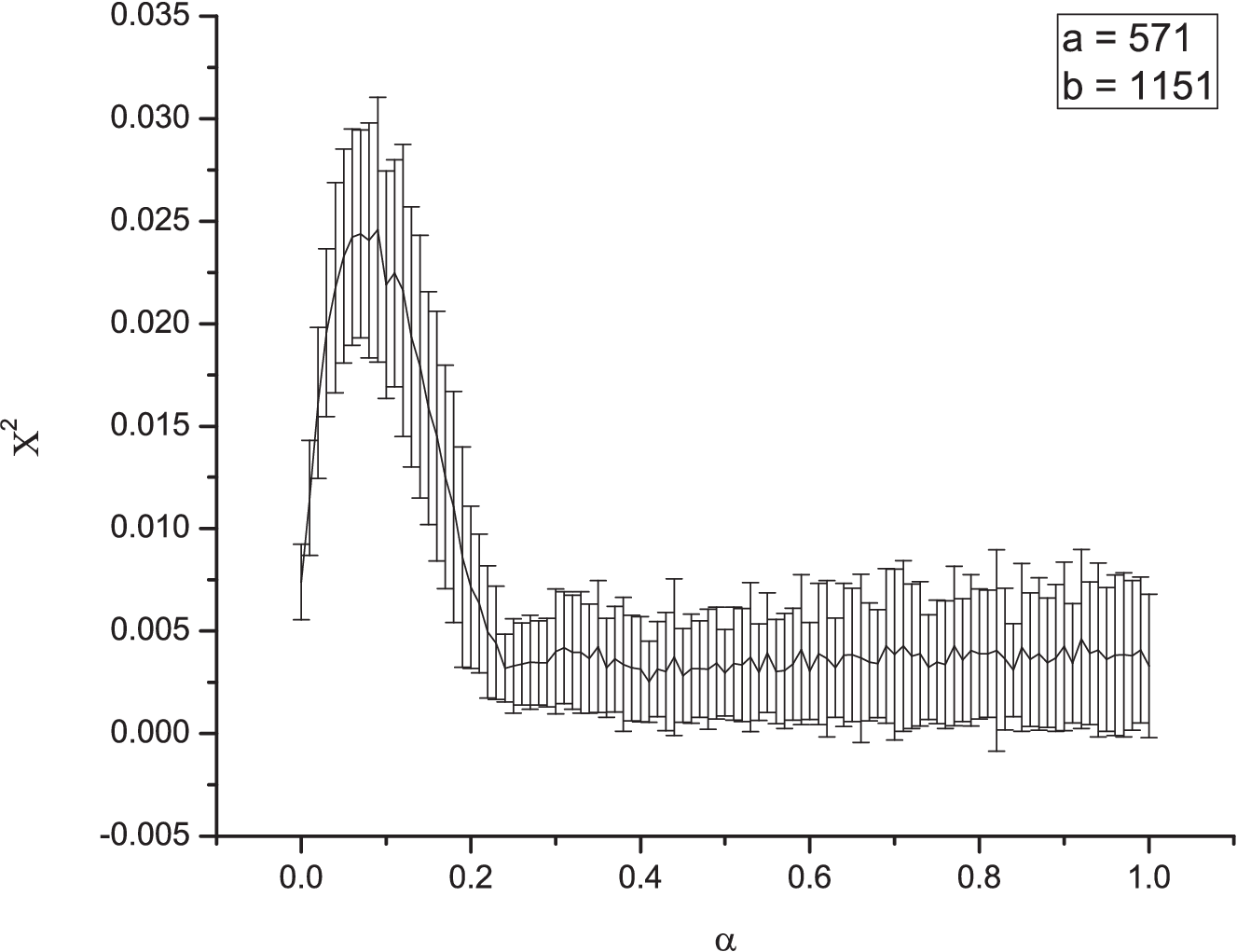}
\caption{The $\chi^2$ for the Kolmogorov's function for the sequence $z_n$ and of its subsequences vs $\alpha$ indicating the contribution of the 
random and regular parts for the pair of indicated input parameters.}%
\label{fig:chi_sq}%
\end{figure}

By fixing the values $a$ and $b$, namely, $a=541$, $b=1151$, we generated 100 different sequences $z_n$ for each value of $\alpha$ within 
$0$ and $1$ by step $0.01$, each sequence containing 10000 elements. Then, each sequence is divided into 50 subsequences and for each subsequence 
the parameter $\Phi(\lambda_n)_m$ is calculated ($m$ runs through values $1, ..., 50$) and the empirical distribution function $G(\Phi)_m$ of these 
numbers was constructed. In the case when the original sequences are random, this distribution should be uniform according to Kolmogorov's theorem. 
Therefore, in general case we  calculated $\chi^2$ of the functions $G(\Phi)_m$ and $G_0(\Phi)=\Phi$ to have an indicator for randomness. Thus one 
parameter $\chi^2$ is calculated per each of the $100\times101$ sequences. Grouping $100$ $\chi^2$ values per one value of $\alpha$, we constructed 
mean and error values for $\chi^2$.
So, for each pair of $a$ and $b$ we get one plot: dependence of $\chi^2$ on $\alpha$.
Two examples of such dependence are given in Fig.\ref{fig:chi_sq}.

Fig.\ref{fig:chi_sq} shows the variation of $\chi^2$ vs $\alpha$ varying from $0$ to $1$, which corresponds to the gradual change of sequences 
$z_n$ from regular to random.

Then we study the behavior of this relation from parameter $a$ for different values of $b$. First we define a parameter $\Delta$ which equals to 
the difference of two values in the above plots: maximal value of $\chi^2$ and minimal value in the range $\alpha \in (0, \alpha_{max})$, 
where $\alpha_{max}$ is the position of the maximal value; e.g. it is obvious that for the left plot $\Delta=0$ and for the right one 
$\Delta \approx 0.02$.

The next step is to fix $b$ and for each value of primary $a={2, ..., b}$ calculate $\Delta$. This is done for several values of $b$.

\begin{figure}%
\includegraphics[width=2.37in]{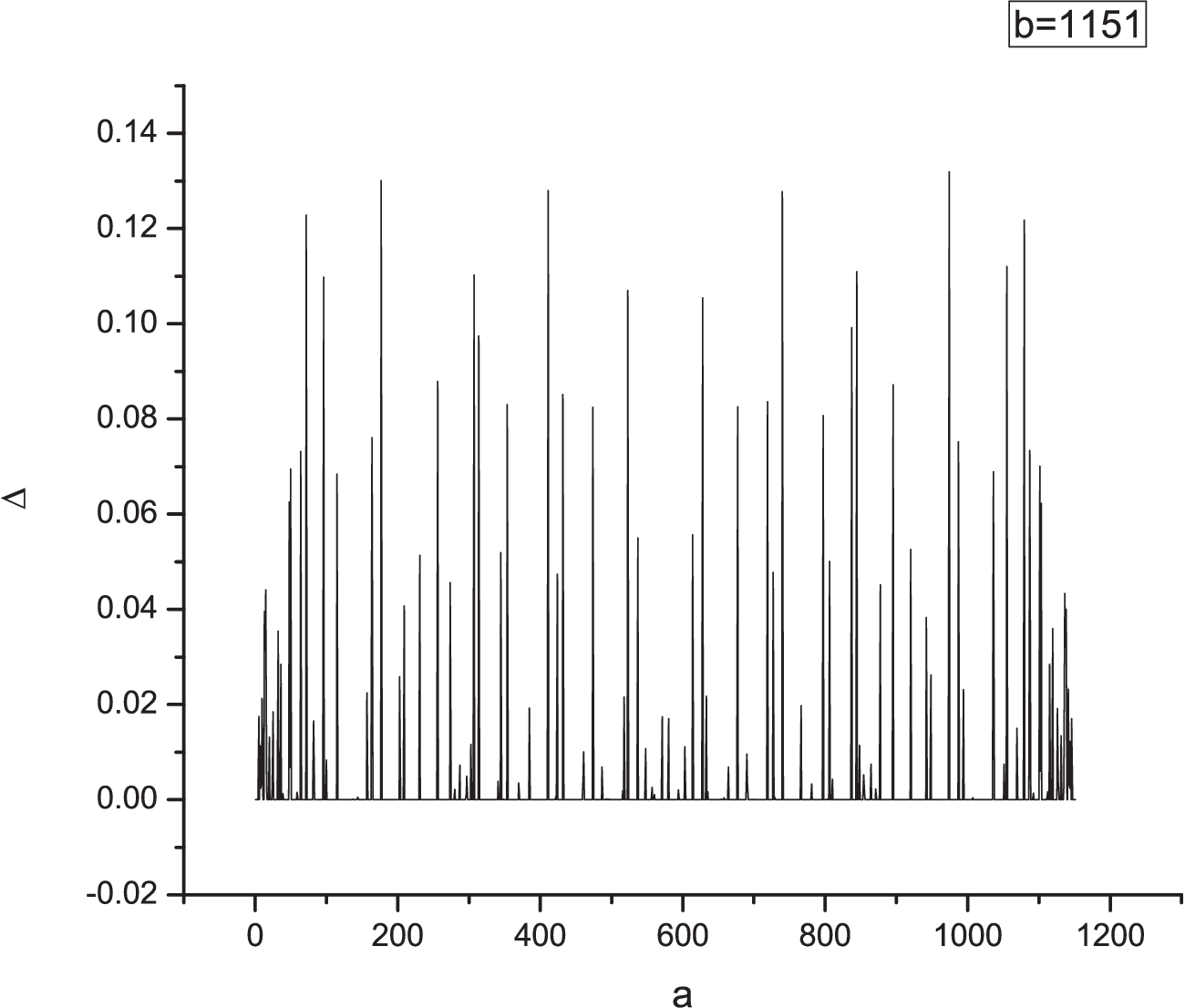}
\includegraphics[width=2.37in]{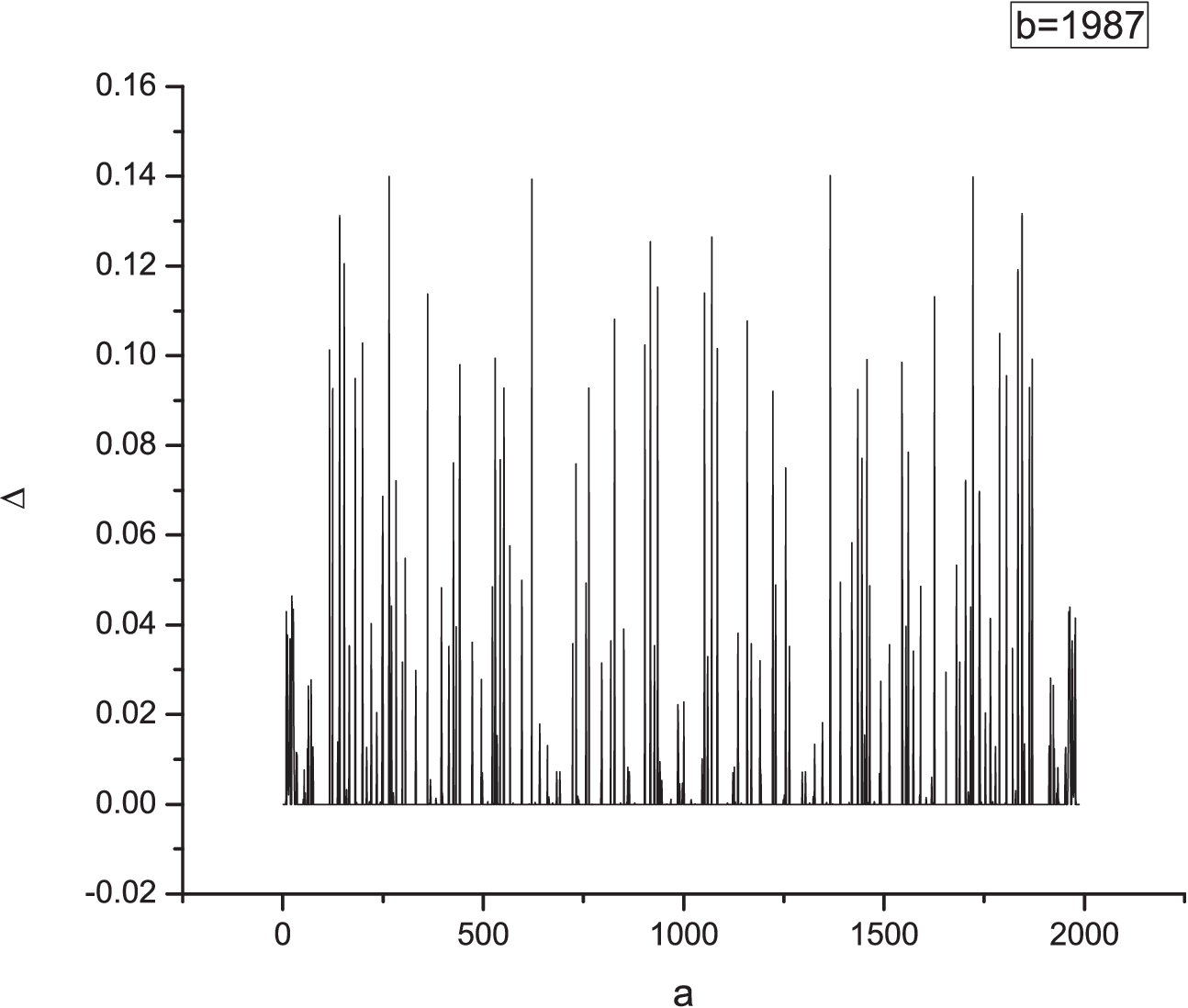}
\caption{The dependence of $\Delta$ vs the parameter $a$.}%
\label{fig:deltas}%
\end{figure}

Fig.\ref{fig:deltas} shows strict mirror symmetry in the dependence of $\Delta$ vs $a$, although no periodicity has been found by Fourier analysis.

Then we proceeded via two types of analysis. In the first, we studied the values and the distribution of $\Delta$ and, in the second, the spacing between 
non-zero $\Delta$ and their distribution.
Since obviously, most values of $\Delta$ are the null ones, we skipped them; also due to the mirror symmetry each $\Delta$ comes with its pair and 
therefore only one of each pair is taken into account. 

\begin{figure}%
\includegraphics[width=2.37in]{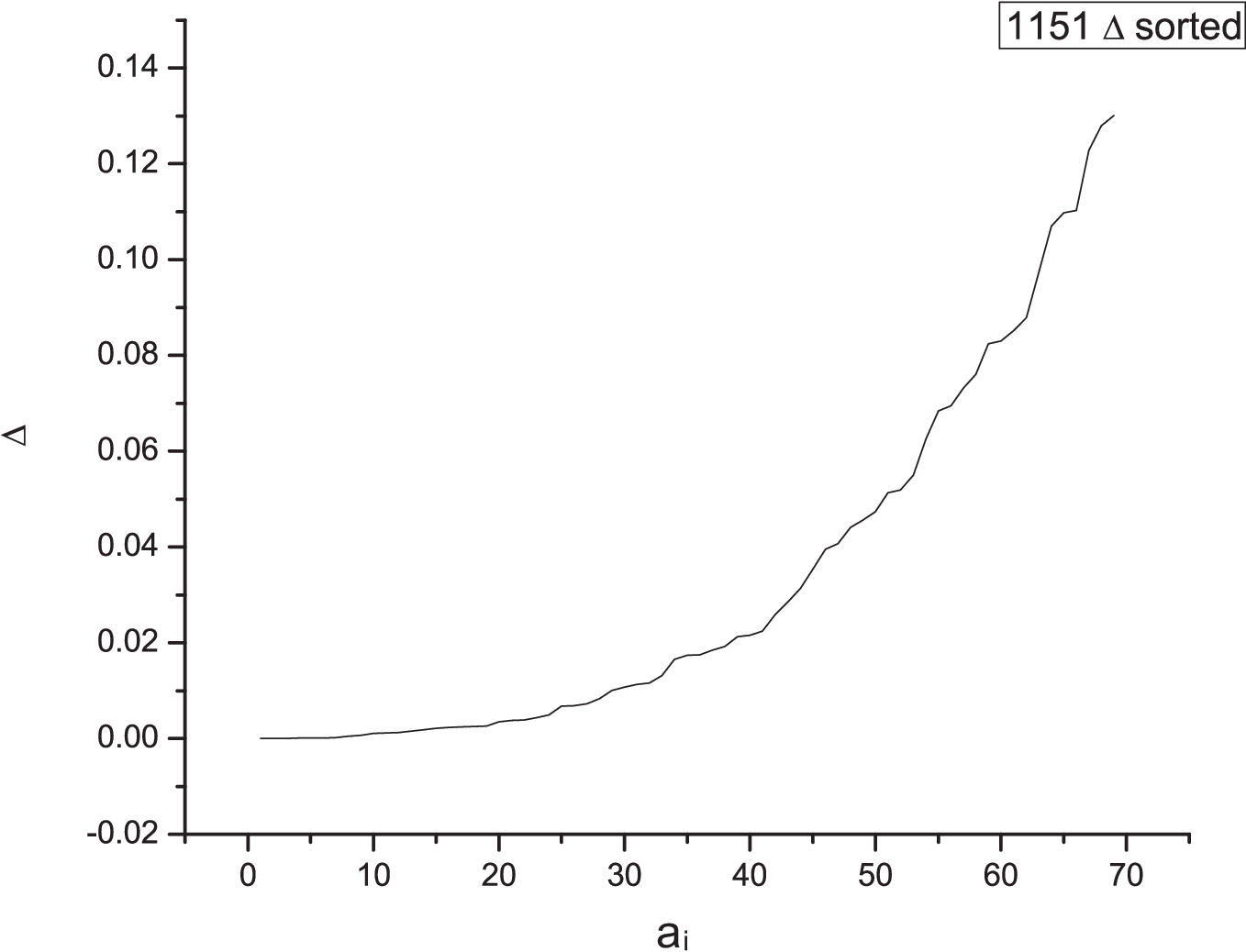}
\includegraphics[width=2.37in]{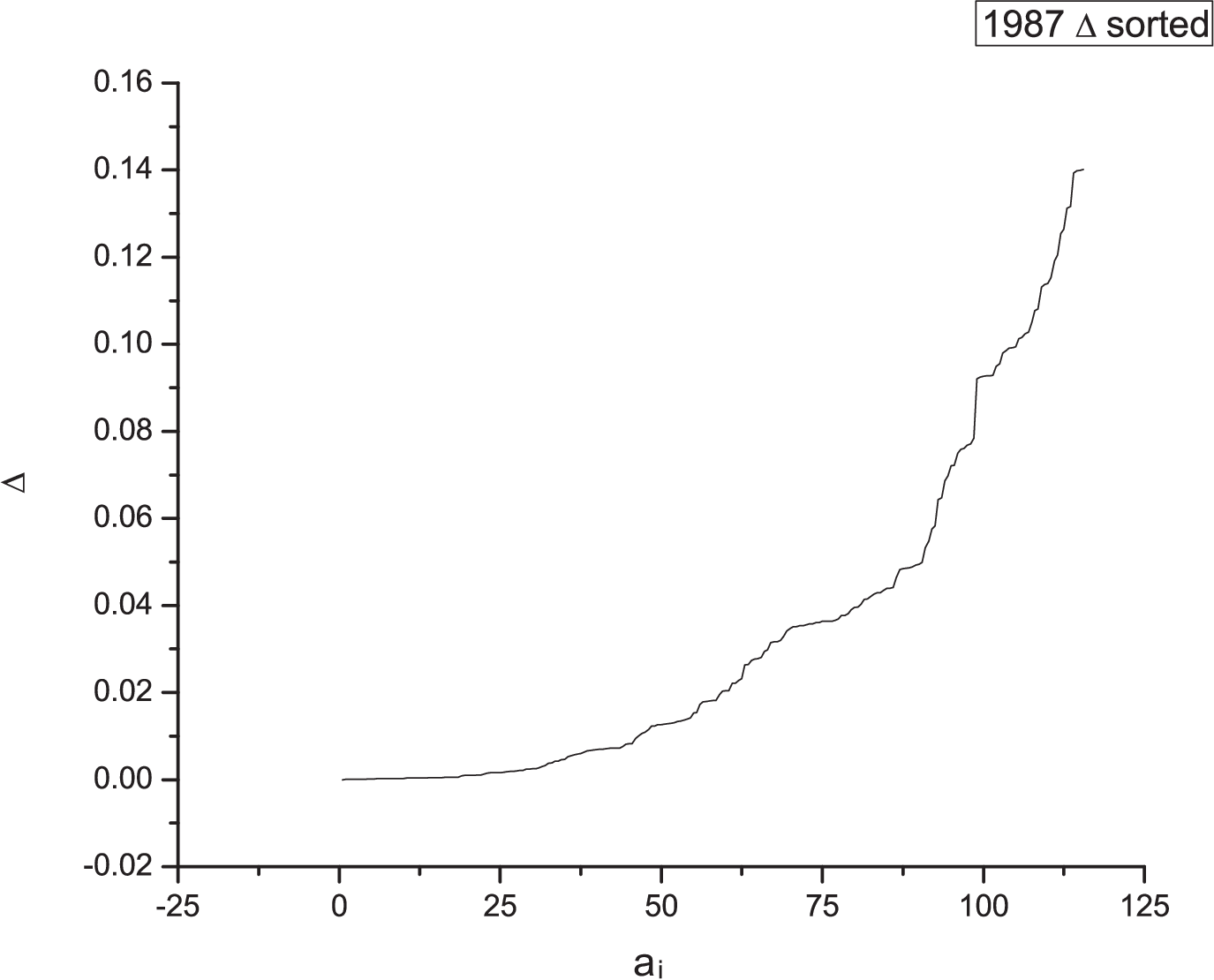}
\caption{Sorted amplitudes of $\Delta$s from Fig.\ref{fig:deltas}.}%
\label{fig:y1}%
\end{figure}

Fig.\ref{fig:y1} exhibits $\Delta$s sorted in growing order, i.e. on the abscissa axis we have the number of non-zero $\Delta$s from Fig.\ref{fig:deltas}. 
The number of non-zero $\Delta$s appear to be proportional to $b$.

\begin{figure}%
\includegraphics[width=2.37in]{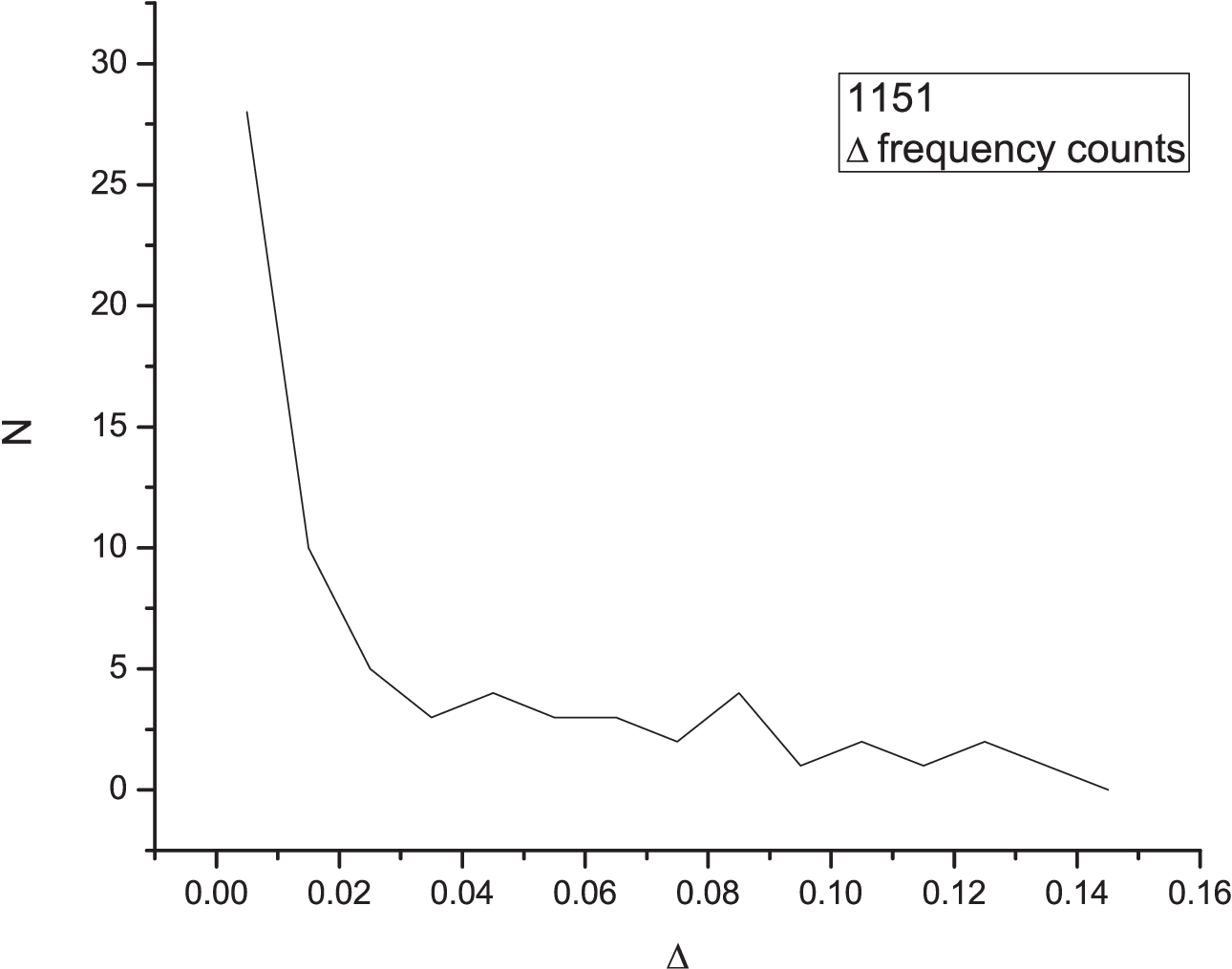}
\includegraphics[width=2.37in]{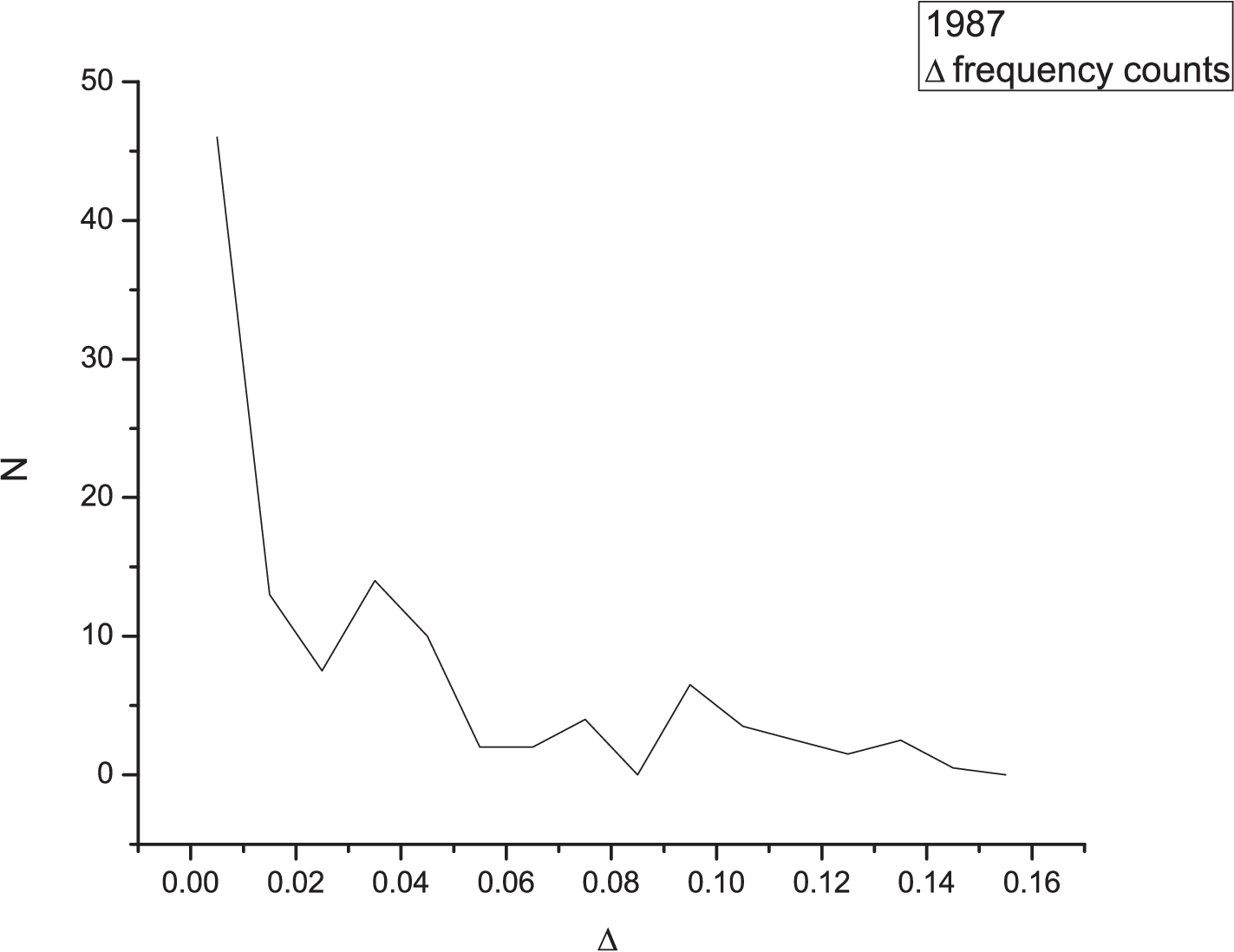}
\caption{The distribution of the amplitudes of $\Delta$ from Fig.\ref{fig:deltas}.}%
\label{fig:y2}%
\end{figure}

Using the same values of $\Delta$s we calculated the frequency counts (Fig.\ref{fig:y2}), which showed the decrease of the number of non-zero 
$\Delta$s with the decrease of their values.

\begin{figure}%
\includegraphics[width=2.37in]{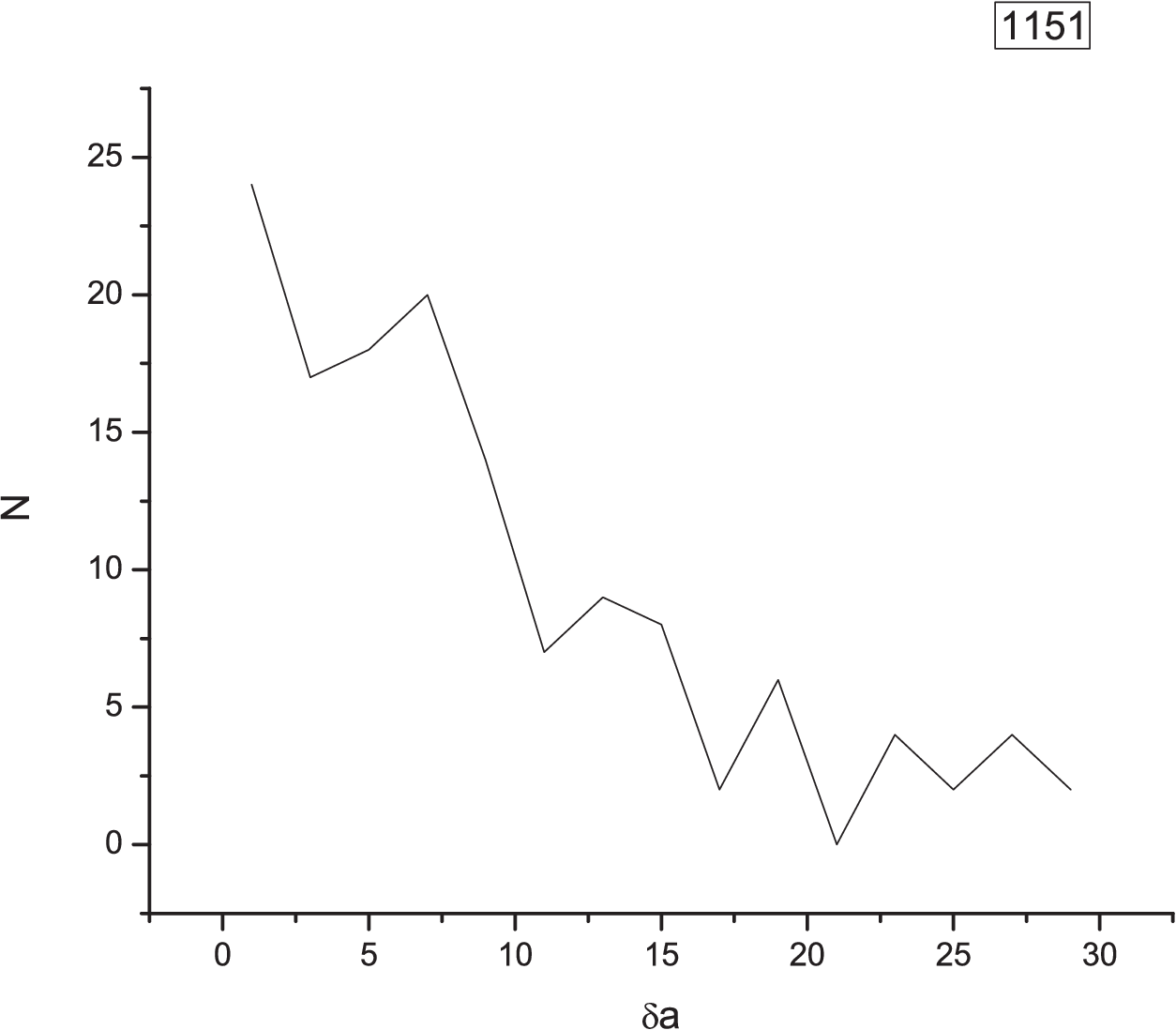}
\includegraphics[width=2.37in]{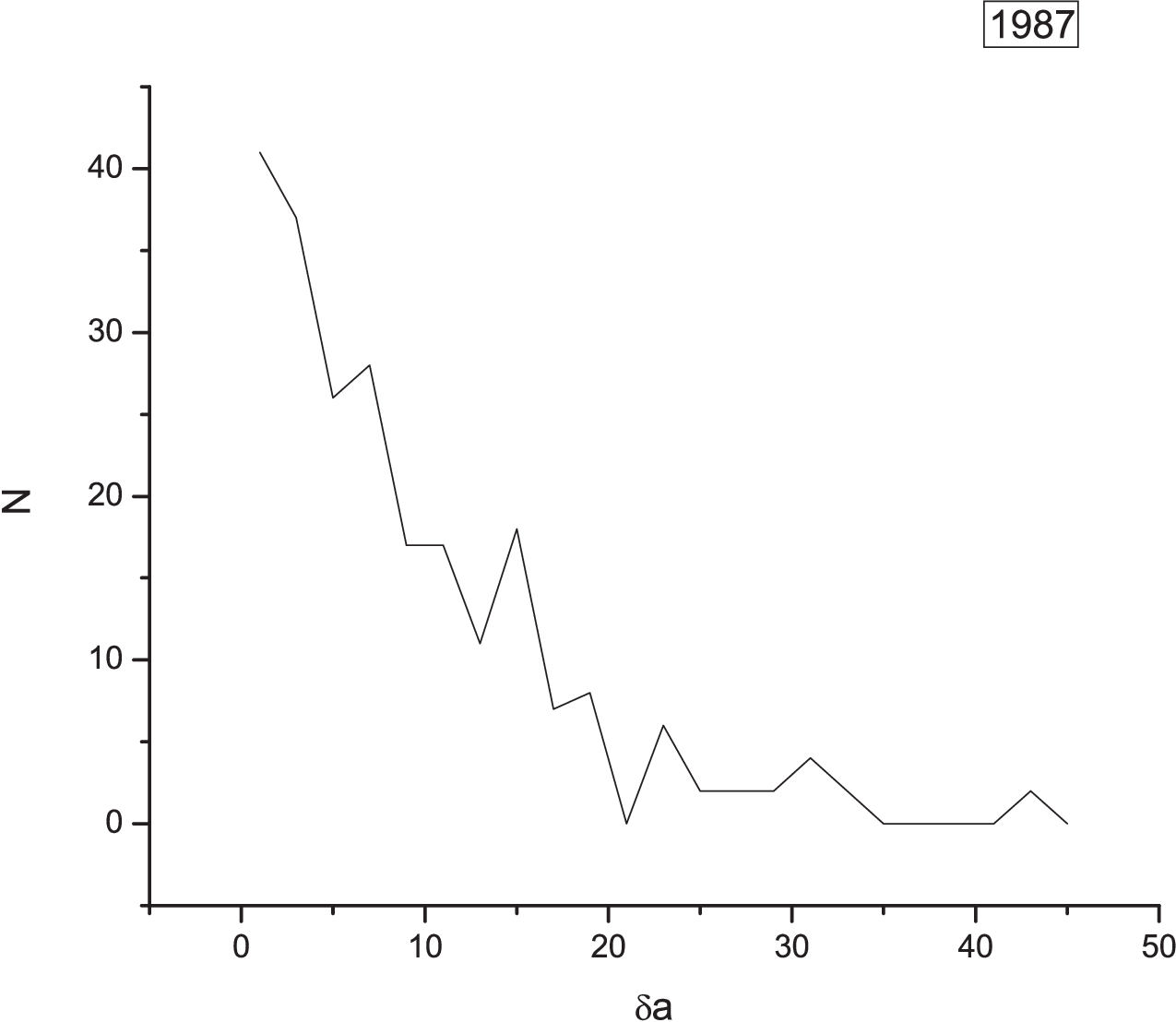}
\caption{The frequency counts for the x-spacing of $\Delta$s.}%
\label{fig:x1}%
\end{figure}

The analysis of the spacings was done in the following way. We collected the distances of all neighbour non-zero $\Delta$s and constructed 
their frequency counts as shown in Fig.\ref{fig:x1}: smaller gaps are more common than larger ones. We also see the maximal gaps between 
non-zero $\Delta$s to be $30-50$, which do not change monotonically depending on $\alpha$.

Now let us follow the features of signals formed as sum of many fluctuations each having the same standard deviation. We generated two type 
of sequences, again of 10000 elements each.

\begin{figure}%
\includegraphics[width=2.37in]{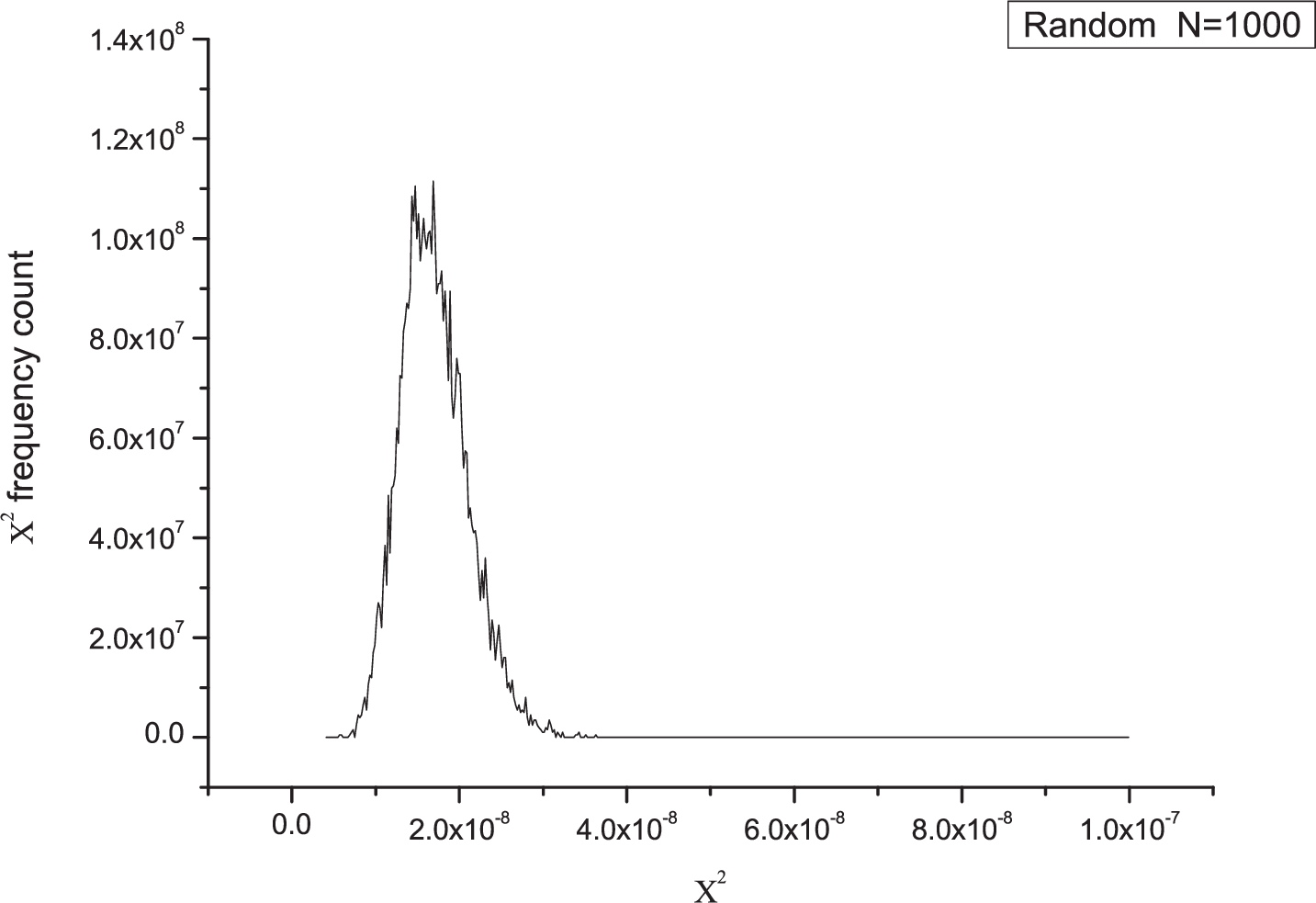}
\includegraphics[width=2.37in]{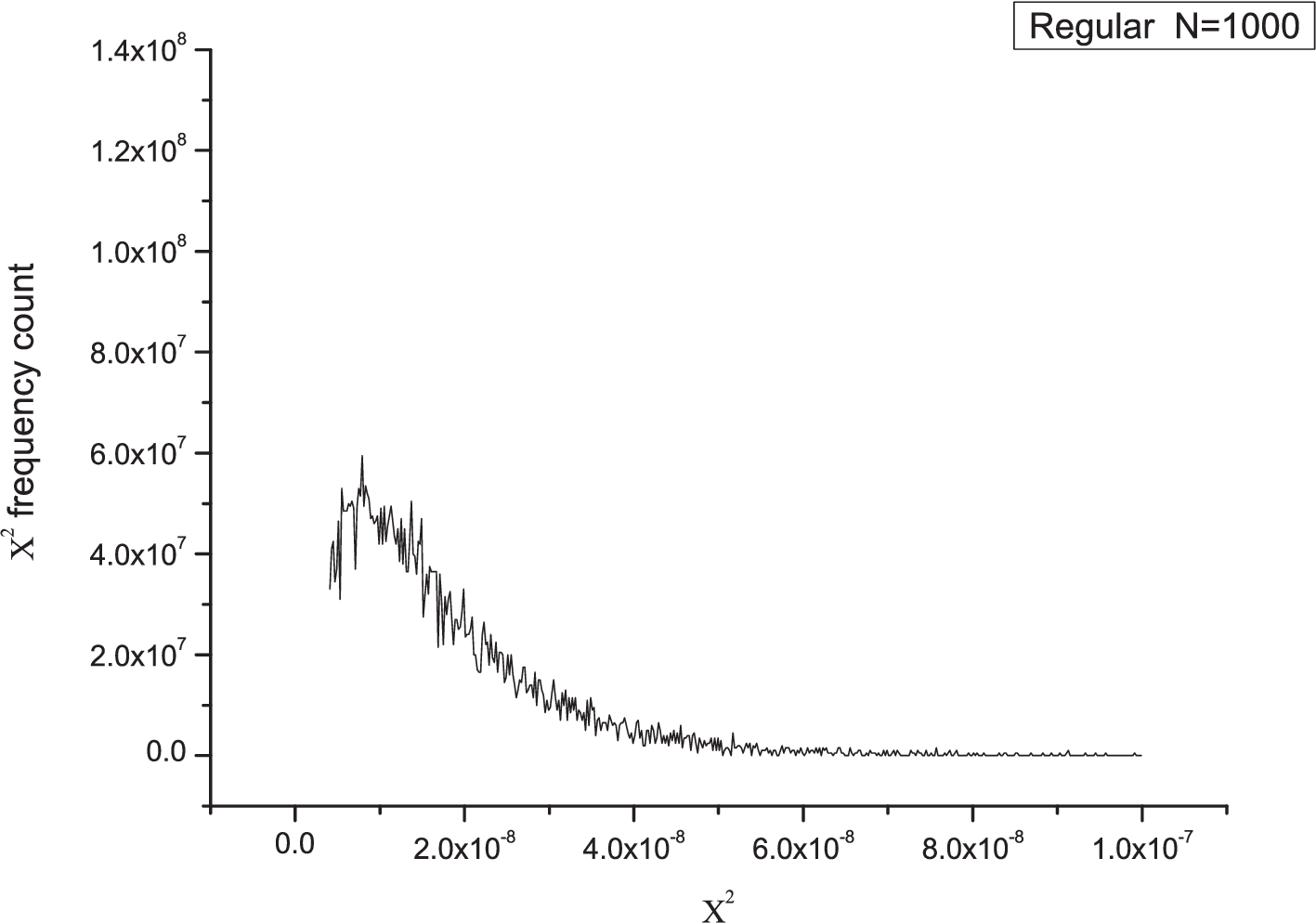}
\caption{$\chi^2$ frequency counts for the set of random and regular sequences as compared
with a Gaussian function.}%
\label{fig:chi_dist}%
\end{figure}

\begin{figure}%
\includegraphics[width=2.37in]{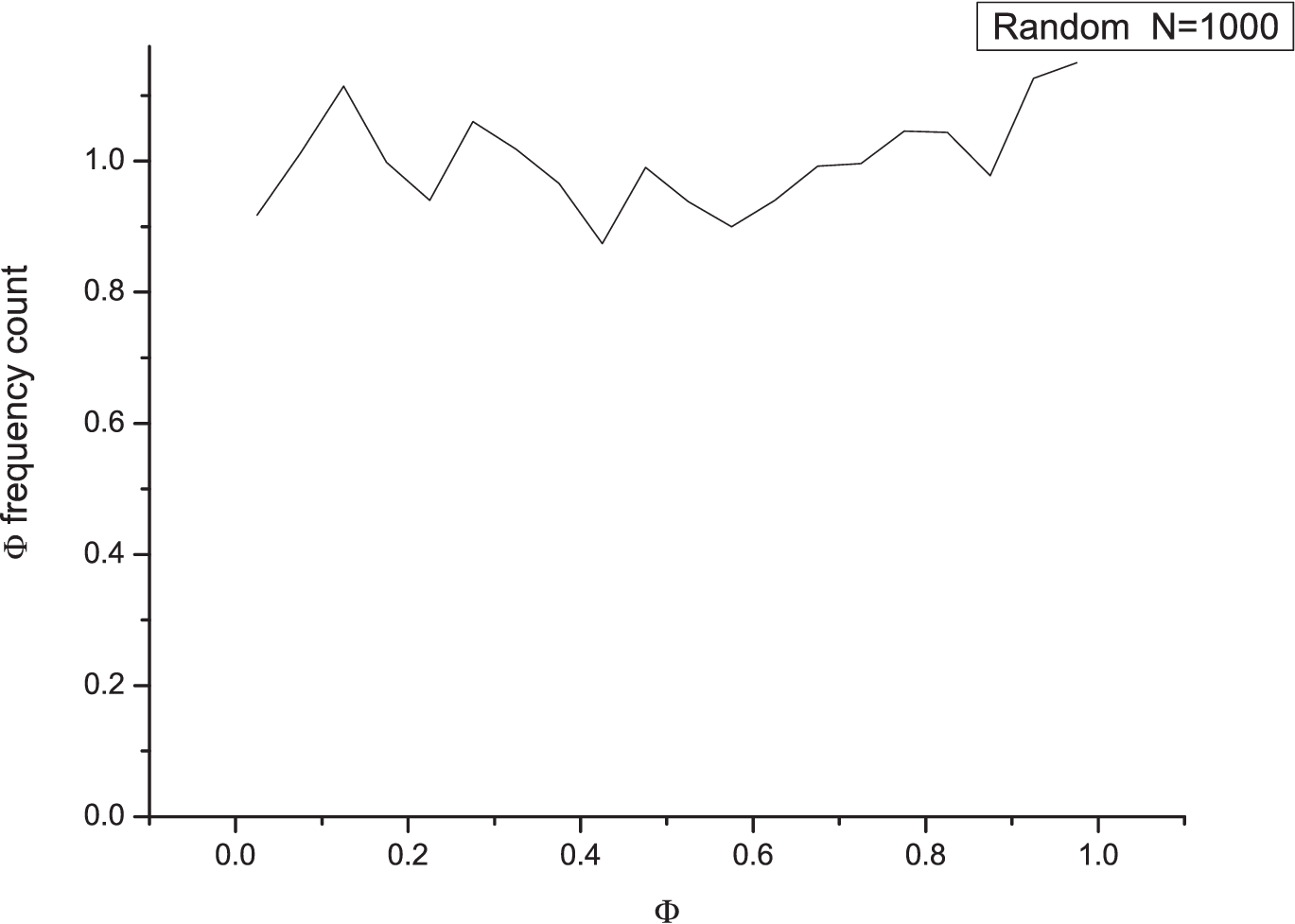}
\includegraphics[width=2.37in]{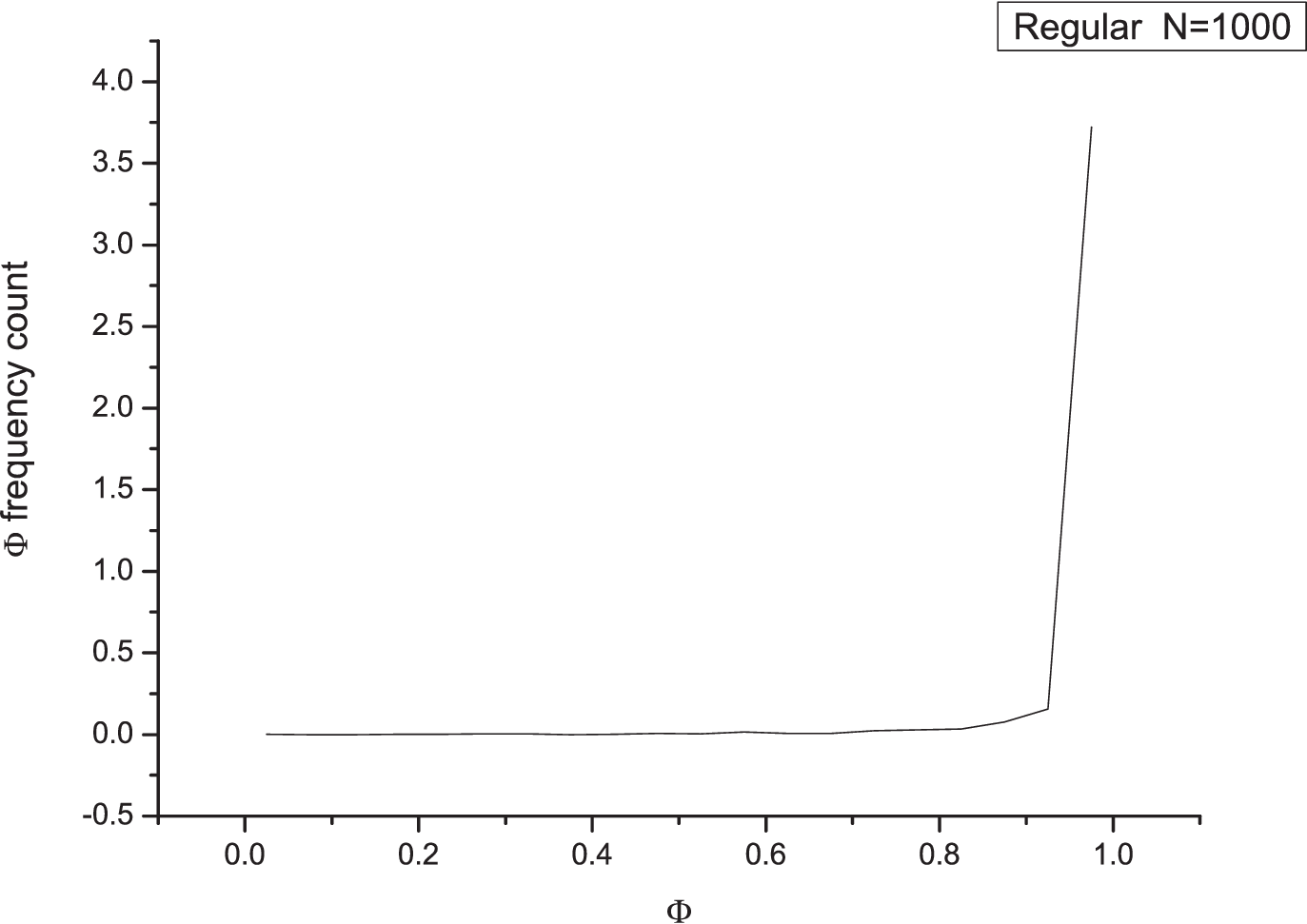}
\caption{Kolmogorov function $\Phi$ for the sequences in Fig.\ref{fig:chi_dist}.}%
\label{fig:phi}%
\end{figure}

\begin{figure}%
\includegraphics[width=3.37in]{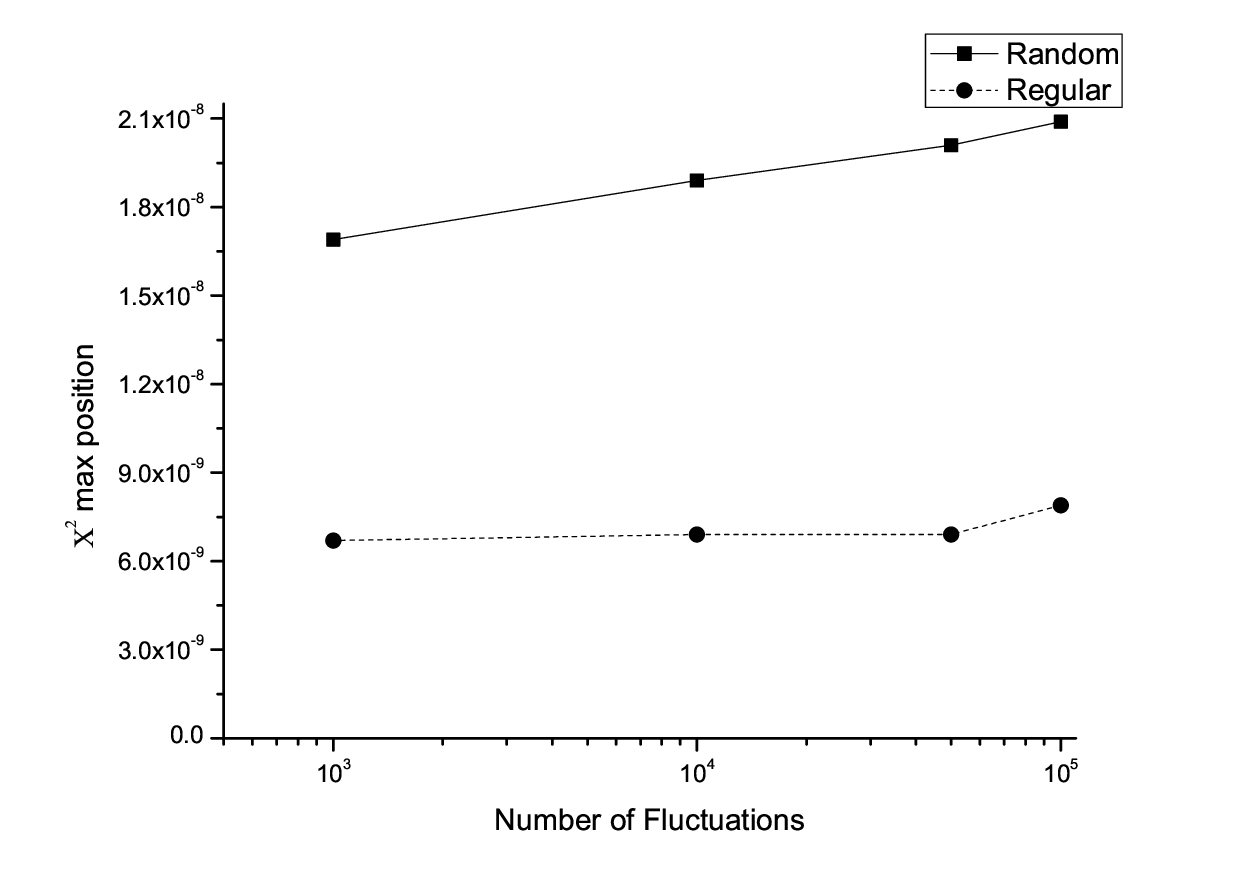}
\caption{The maxima of the $\chi^2$ in Fig.7 vs the number of the fluctuations.}%
\label{fig:N}%
\end{figure}

The first one is given as
\begin{equation}
a_i=\frac{1}{\sqrt N} \sum_{k=1}^N Compact(x_i^k,-1,1),
\end{equation}
where $x_i^k=i/k$, i.e. $x^k$ is compactified arithmetical sequence within the 
interval $(-1,1)$, with step $1/k$. We call this a regular sequence.

The second sequence is
$$
b_i=\frac{1}{\sqrt N} \sum_{k=1}^N Random(-1,1),
$$ 
which we denote as a random one. 

Here we used the following notation: $Compact(x,p,q)=q+x mod(p-q)$ indicates multiples of $(p-q)$ from $x$ having the value within the 
range $(p,q), p<q$. The comparison of the sequences is done by means of Kolmogorov parameter with varying also the values of N.

The results for 10000 random (generated by random number generator) and regular sequences each in Fig.\ref{fig:chi_dist} show the typical differences for $\chi^2$ 
when the number of the fluctuations vary from $N=1000$ to $100000$.
The equal scales of $\chi^2$ shows that in both cases i.e. random and regular sequences, we deal with a Guassian
distribution, in accordance with the Central Limit theorem. That theorem states that for large enough values of $N$, both sequences $a_i$ and 
$b_i$ tend to Gaussian sequences with the same $\sigma$ and $\mu$ not depending on $N$. Along with that, however, we see differences in the 
Gaussians, namely, the standard deviations are larger for the regular case. 

Involving then the Kolmogorov function $\Phi$, we see that although we have Gaussians both in random and regular cases, the behavior of $\Phi$ 
is quite different, as shown in Fig.\ref{fig:phi}, i.e. it is close to a homogeneous function for random sequences and $\Phi=1$ for regular ones. 
So, the Kolmogorov's function enables to distinguish the superposition of random and regular sequences, although both are tending to Gaussians.  
  
This analysis does not show any significant dependence on the length of the sequences, also dependence on the number of the fluctuations within 
$1000-100000$ is rather weak as seen in Fig.\ref{fig:N}. 

\section{Conclusions}

The performed numerical analysis aimed to reveal the behavior of the Kolmogorov distribution vs the randomness of generated signals. To describe 
astrophysical datasets which contain both regular and stochastic components, we considered sequences scaled by a single parameter $\alpha$, the 
ratio of those components. 

Both qualitative signatures and their quantitative scalings have been observed at the numerical experiments. 
Namely, the appearance of a critical value of $\alpha$ has been shown, which defines the qualitative change in the behavior of the frequency count 
of the Kolmogorov distribution: the monotonic decay is transformed to a function with extremum. Then, the dependence of these features vs the 
$\alpha$ reveals mirror properties in the amplitude and distribution of the frequency counts of the function $\Phi$. Then, the behavior of the 
randomness for large number of subsignals has been also revealed, where the Kolmogorov function acts as an informative descriptor. Particularly, 
the descriptor is informative at large $N$ limit when although the sum both of random and regular signals tends to a Gaussian in accordance to the Central Limit theorem, the set of regular subsignals is distinguished by a notably larger variance. In the case of the cosmic microwave backgound, for example, the contributions to the overall detected signal, besides the cosmological one, typically are due to the Galactic synchrotron and dust emissions, star forming galaxies, point sources (quasars, blazars), instrumental noise, etc.   

The behaviors obtained at the numerical experiments due to the universality of the approach will serve as indicators in the study of various 
astrophysical signals, including in revealing the contributions of correlated and random components.
      
We are thankful to A.A.Kocharyan for many useful discussions.

\end{document}